# Exploring Non-Linear Effects of Built Environment on Travel Using an Integrated Machine Learning and Inferential Modeling Approach: A Three-Wave Repeated Cross-Sectional Study

**Niaz Mahmud Zafri**
Doctoral Student
Department of Urban Studies and Planning
Massachusetts Institute of Technology
Cambridge, MA, 02139, United States
Email: zafri@mit.edu

**Ming Zhang**
Professor
School of Architecture
University of Texas at Austin
Austin, TX, 78712, United States
Email: zhangm@austin.utexas.edu

**Corresponding Author:** Niaz Mahmud Zafri

# Exploring Non-Linear Effects of Built Environment on Travel Using an Integrated Machine Learning and Inferential Modeling Approach: A Three-Wave Repeated Cross-Sectional Study

## Abstract

This study investigates the dynamic relationship between the built environment and travel in Austin, Texas, over a 20-year period. Using three waves of household travel surveys from 1997, 2006, and 2017, the research employs a repeated cross-sectional approach to address the limitations of traditional longitudinal and cross-sectional studies. Methodologically, it integrates machine learning and inferential modeling to uncover non-linear relationships and threshold effects of built environment characteristics on travel. Findings reveal that the built environment serves as a sustainable tool for managing travel in the long term, contributing 50% or more to the total feature importance in predicting individual travel—surpassing the combined effects of personal and household characteristics. Increased transit accessibility, local and regional destination accessibility, population and employment density, and diversity significantly reduce travel, particularly within their identified thresholds, though the magnitude of their influence varies across time periods. These findings highlight the potential of smart growth policies—such as expanding transit accessibility, promoting high-density and mixed-use development, and discouraging single-use development and peripheral sprawl—as effective strategies to reduce car dependency and manage travel demand.

# 1. Background of the Study

In the era of sustainable development, cities worldwide face challenges due to excessive car use. Despite goals to reduce car dependency, car usage continues to rise. To address this, research emphasizes reducing car use through concepts like new urbanism and smart growth, which promote high-density, compact, mixed-use, walkable, and transit-oriented development (Ewing & Cervero, 2010; M. Zhang, 2004). These approaches aim to reduce trip distances and encourage transit and active transportation, potentially decreasing car dependency and its associated issues, such as congestion, pollution, and inequity.

The relationship between the built environment and travel behavior has long interested researchers and practitioners in transportation, urban planning and design, health, and other fields. Research in the past three decades has accumulated voluminous publications on the subject matter. While a good portion of the existing literature concluded that built environment could influence travel behavior, the reported findings on the magnitude of influence of the built environment on travel outcomes remain mixed. A meta-analysis of the existing literature by Stevens (2017) indicated that the impact of compact development on reducing driving is marginal, sparking debate among scholars (Clifton, 2017; Ewing & Cervero, 2017; Handy, 2017; Heres & Niemeier, 2017; Knaap et al., 2017; Manville, 2017; Nelson, 2017). Some of the key issues identified through this debate include the reliance on linear models and cross-sectional data, which may have underestimated the impact of the built environment on travel behavior in earlier studies (Clifton, 2017; Handy, 2017). Consequently, recent research has increasingly focused on exploring the non-linear nature of this relationship and utilizing data beyond a cross-sectional framework. The current state of knowledge, challenges, and potential directions for research are discussed below.

## 1.1 Studies on non-linear relationship

Traditional inferential models have long served as the cornerstone for examining the relationship between the built environment and travel behavior. These models offer valuable insights into the magnitude and statistical significance of these relationships, providing interpretable coefficients and elasticities that are essential for policymaking. However, the assumption of linearity often oversimplifies the complex interactions between built environment factors and travel behavior, overlooking potential non-linear effects and threshold behaviors (Aghaabbasi & Chalermpong, 2023; Clifton, 2017; van Wee & Handy, 2016; Wu et al., 2019). Consequently, reliance on linear modeling can lead to inaccurate estimates, underestimating the true impact of built environment factors.

While piecewise regression models have the ability to address non-linearity, they depend heavily on researcher-specified breakpoints, often determined through trial-and-error methods. This approach introduces the risk of bias, as the choice of breakpoints can be influenced by assumptions that may not accurately reflect the underlying data structure.

In response to these limitations, recent studies have increasingly focused on the non-linear and threshold effects of the built environment on travel behavior, including mode choice (Ashik et al., 2024; Ding, Cao, & Wang, 2018; Hatami et al., 2023), driving distance (Ding, Cao, & Næss, 2018), driving (Hu et al., 2021; W. Zhang et al., 2022a), and emissions (Shao et al., 2023; Wu et al., 2019). Machine learning methods have been widely applied to explore non-linearity and threshold effects (Aghaabbasi & Chalermpong, 2023; Ashik et al., 2024; Hatami et al., 2023; Hu et al., 2021; Shao et al., 2023). While machine learning can efficiently uncover non-linear relationships and threshold effects without relying on a priori assumptions, it lacks the ability to quantify the magnitude of these relationships or provide measures of statistical significance, which limits its explanatory power and policy relevance (Ding, Cao, & Wang, 2018; Shao et al., 2023; Wu et al., 2019). Moreover, distinguishing between true non-linear effects and spurious patterns caused by random noise or chance correlations remains a challenge. The focus on single time periods or dataset further raises concerns about the stability and consistency of the non-linear results across time or beyond specific dataset (X. Wang & Cheng, 2020).

By integrating machine learning with traditional inferential modeling, it is possible to bridge these gaps, leveraging the strengths of both approaches. Machine learning can initially be employed to identify non-linear relationships and thresholds among built environment factors. These findings are then incorporated into traditional models through piecewise regression, which provides elasticity estimates and statistical significance for the identified relationships. This dual-method approach not only enhances the robustness of the analysis but also improves interpretability, allowing for more nuanced and actionable insights into the built environment–travel relationship.

To ensure the stability and consistency of these findings, it is crucial to study non-linear relationships across multiple time periods and datasets. If such relationships remain stable over time and across multiple datasets, this increases confidence in their validity and reduces the likelihood that they result from random noise or dataset-specific anomalies. Repeated cross-sectional studies spanning extended periods and datasets are essential in this regard, as they not only capture long-term trends but also help assess whether identified non-linear relationships persist across different time frames and datasets (M. Wang et al., 2017). Such studies provide deeper insights into the enduring dynamics of the built environment and travel behavior, ensuring that policy recommendations are grounded in stable and well-supported evidence.

## 1.2 Studies on multi-period data

Cross-sectional studies provide valuable insights into the correlational relationship between the built environment and travel behavior at specific points in time. However, they are limited in establishing causality and accurately quantifying impact magnitude (Clifton, 2017; Coevering et al., 2016; Handy et al., 2005). Studies using multi-period data (longitudinal data) can address these limitations by offering a more comprehensive view over time, though such data are challenging to obtain, particularly at a disaggregated level. Some studies have utilized multi-period data, which can generally be categorized into two types: a) panel data studies and b) household relocation studies.

A few disaggregated studies have attempted to conduct longitudinal research by utilizing panel data (Coevering et al., 2016; Gao et al., 2019; Kamruzzaman et al., 2016; Rahman, 2023; van de Coevering et al., 2021). These studies use panel survey data from the same samples across multiple periods (generally two waves), usually with short intervals (less than 10 years), to estimate the association between certain aspects of the built environment and travel behavior, while controlling for factors like travel attitudes and socio-demographics. However, panel data studies face limitations, such as the overrepresentation of wealthy and educated samples, stagnation effects, and the relative stability of built environment and travel behavior over short periods in developed countries. They also struggle to account for changes in travel behavior due to life events and other external variables, leading to potential inaccuracies. While panel data can help establish causality, these studies provide limited insights into the magnitude of relationships due to the frequent use of structural equation modeling (SEM) techniques and other abovementioned limitations.

On the other side, some studies focus on household relocation and use quasi-longitudinal data from individuals who have recently moved (Cao et al., 2007; De Vos et al., 2018; De Vos, Cheng, Kamruzzaman, et al., 2021; De Vos, Cheng, & Witlox, 2021; Handy et al., 2005, 2006; Krizek, 2000, 2003; Milakis et al., 2017; D. Wang & Lin, 2019; Wells & Yang, 2008). These studies assume that changes in residential location also lead to changes in the built environment and, subsequently, affect travel behavior. However, these studies are retrospective, relying on memory and using imprecise indicators, which could affect reliability and accuracy of the findings. While they help in understanding causal relationships, they often struggle to quantify the magnitude of the built environment's effects on travel behavior due to the imprecise nature of the data and the often use of SEM-based analysis methods.

Besides these two types of studies, a third type of longitudinal study is rarely found in literature. This type, known as a repeated cross-sectional study, is a "pseudo-longitudinal" type of data that combines data from two cross-sectional household travel surveys to create a longitudinal dataset (Grunfelder & Nielsen, 2012; Zhang & Zhang, 2015). Although this type of study can estimate the magnitude of impacts more accurately, previous studies typically cover a short period (less than 10 years). Generally, built environment characteristics and travel behavior remain stable over such short periods in developed cities.

In summary, while longitudinal studies help establish causal relationships, they often fail to provide reliable evidence on the magnitude of impacts due to inherent limitations. Therefore, studies with more than two waves of data over a longer timeframe (around 20 years) in rapidly growing cities, where significant changes in the built environment occur, are needed to accurately assess the magnitude of the built environment's impact on travel behavior.

## 1.3 Research objectives and contributions

The primary objective of this study is to deepen the understanding of the relationship between the built environment and travel (vehicle miles traveled (VMT)) in Austin, TX, over a 20-year period. Utilizing three waves of household travel surveys from 1997, 2006, and 2017, this research uses a repeated cross-sectional approach to analyze the dynamics of the magnitude of built environment and travel connection. Additionally, the study integrates machine learning and traditional inferential modeling to explore non-linear relationships and identify threshold effects over time. Through this approach, this study aims to answer two main questions: (a) Can the built environment be sustained as a tool for controlling travel in the long term? and (b) How have the built environment and travel relationship and its magnitude evolved over the considered time?

Our study area, Austin, has experienced rapid growth and significant changes over the past three decades, with the population and employment nearly doubling. Notable changes include downtown growth, neighborhood infill and redevelopment, fringe expansion, and the expansion of the public transit network. Additionally, transit-oriented development and smart growth initiatives emerged after 2005. Overall, the extensive transformation of the built environment in Austin makes this study area ideal for such studies.

This study contributes significantly to the existing literature by uniquely integrating machine learning and traditional inferential modeling to explore non-linear relationships and identify threshold effects over a 20-year period, which clarifies the true nature of the relationship and leads to more reliable findings. Additionally, this study also contributes to deepening the understanding of the relationship by adopting a repeated cross-sectional approach in a rapidly growing city and highlighting the magnitude of the impact over time. Lastly, by focusing on VMT as a measure of travel behavior, the study provides valuable insights into how built environment factors influence travel demand and distance in the context of a U.S. city.

## 2. Data

### 2.1 Data preparation

To conduct this study, we collected data of three household travel surveys conducted in 1997, 2006, and 2017 in the Austin, Texas region from the Capital Area Metropolitan Planning Organization (CAMPO). These surveys covered a large, randomly selected number of households within the Austin region and collected all trip data for all household members during a workday. Although the study area coverage varied across the three surveys, all three consistently included the counties of Travis, Williamson, and Hays. To maintain consistency across the three time periods, we only considered household data from these three counties. A map of the study is presented in **Figure 1**.

In this study, our dependent variable is individual daily VMT. For the 2006 and 2017 travel surveys, origin and destination coordinates were available for all trips. However, for the 1997 data, precise origin and destination coordinates were not provided. Instead, traffic analysis zone (TAZ) IDs for the origins and destinations of each trip were available. Using the spatial file for the 1997 TAZs, we estimated the coordinates of TAZs based on the centroids of these zones. These estimated coordinates were then used as coordinates for the origin and destination of the corresponding trip.

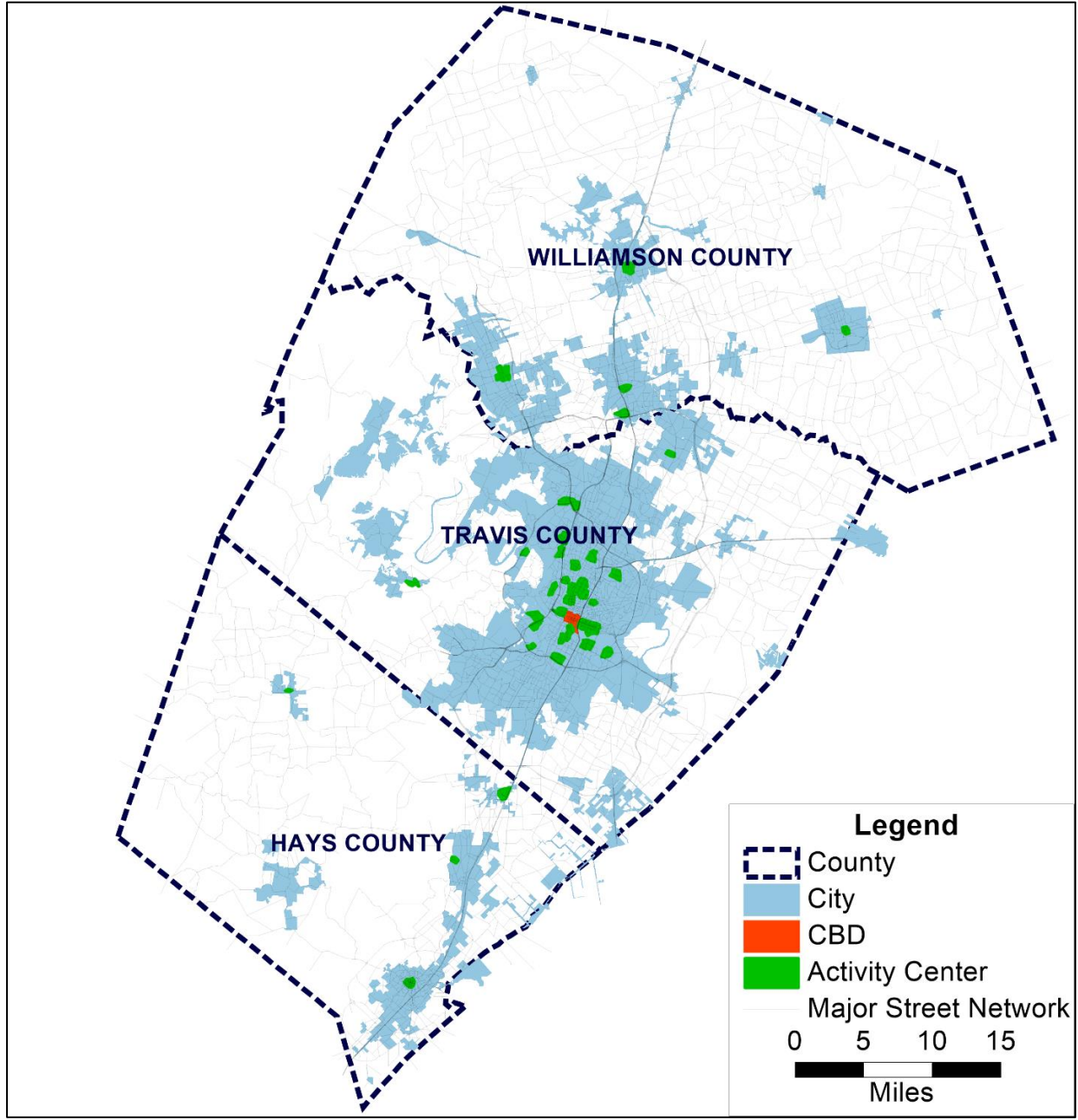

**Figure 1**: Map of the study area

Using data of the trip coordinates, we estimated person miles traveled (PMT) for all trips using the shortest route distance technique in TransCAD software. From PMT, we estimated VMT using the mode of transportation information. If a person made a trip using any personal motorized vehicle, such as a car, motorcycle, van, jeep, etc., the PMT was considered the VMT for that person. We then aggregated the VMT of all trips for each person to estimate their total daily VMT, which is the dependent variable of our study. We removed data of an individual from the sample if they traveled outside the study area, as accurate origin and destination coordinates or information for external areas were unavailable in the surveys, and road network information outside the study area was not available. Only a small number of records (~5% of the overall sample) were removed through this process. Finally, we processed data for 4459, 3020, and 6252 individuals from 1854, 1128, and 2459 households for the years 1997, 2006, and 2017, respectively.

In the case of explanatory variables, data related to personal and household characteristics for each individual were available in the household travel surveys. Summary statistics of this data are presented in **Table 1**. Residential self-selection is a widely discussed factor in the literature that must be controlled to address endogeneity and accurately estimate the built environment's impact on travel behavior (Ewing & Cervero, 2010; D. Wang & Lin, 2019; M. Zhang & Zhang, 2020). Following the method of M. Zhang & Zhang (2020) and W. Zhang & Zhang (2015), we aimed to control the effect of residential self-selection using a variable named HH_Selection, which indicates the factors influencing the respondent's current household location choice. This variable is classified as an access factor if the household location choice is influenced by proximity to work, school, or public transportation; otherwise, it is considered a non-access factor.

For built environment characteristics data, previous studies have considered a wide range of indicators, especially those representing the 5Ds: density, diversity, design, distance to transit, and destination accessibility (Ewing & Cervero, 2010; Handy, 2017; Nelson, 2017). Since we needed consistent data for all three years, we faced limitations in selecting a wide variety of built environment factors. We had TAZ spatial boundaries and corresponding population and employment data for all three years, which were collected from the CAMPO. Therefore, we were able to incorporate TAZ-level population density and employment density for all three years.

We were unable to collect land use data for all three years for the entire study area. So, we could not include land use diversity. However, we did have disaggregated TAZ-level employment data for basic, retail, and service sectors, as well as household data. Using this information, we estimated TAZ-level 3-tier employment (basic, retail, and service) and household entropy. This was considered the diversity variable in this study. The equation for diversity calculation is presented in **Table 1**. Then, based on the geographic coordinates of each individual's household location, we identified the corresponding TAZ. We then attached TAZ-level population density, employment density, and diversity data to each corresponding individual.

In addition, we incorporated distance to transit related built environment characteristics by estimating the distance to the nearest transit stop from each individual's household location. Since we were unable to collect transit stop data for 1997, we estimated this indicator based on transit stop location data from 2003, assuming that the number and locations of transit stops were relatively stable between 1997 and 2003. Although this assumption may not be perfectly precise, it is reasonable given that the transit network in Austin expanded significantly after 2006. Therefore, we do not expect this to cause significant errors in the modeling process.

Additionally, we considered distance to the Central Business District (CBD) and distance to the nearest activity center to address the regional and local destination accessibility dimension of the built environment, respectively. Austin is a monocentric city with a single downtown area. Thus, the distance to the CBD was estimated as the distance from each household location to downtown. In contrast, in 2005, the City of Austin identified activity centers in the Austin region. Using these locations, we estimated the distance from each household location to the nearest activity center. In this way, we tried to minimize data limitations and incorporated relevant built environment characteristics to achieve a comprehensive and robust result. Summary statistics of built environment characteristics are presented in **Table 1**.

## 2.2 Sample Characteristics

Based on the sample characteristics provided in **Table 1**, the average daily VMT of individuals was highest in 1997 and significantly decreased by 2006. In 2017, the average daily VMT was slightly lower than in 2006. This reduction in VMT may be attributed to different age distributions in the samples across the three years, along with other factors. The 1997 sample was considerably younger than those in 2006 and 2017. Additionally, the 1997 sample had a higher proportion of individuals in the working group who were less likely to work from home compared to the samples from 2006 and 2017. Other variables related to personal characteristics exhibited roughly similar distributions across the three time periods.

Regarding household characteristics, the number of motorized vehicles and bicycles in households increased over time. In 1997, only 13% of the sample chose their household location based on access-related factors. This percentage increased to 28% in 2006 and 33% in 2017, indicating that more people chose their household location based on access-related factors in the later years. Household income was also higher in the later years, which may indicate higher-income samples in those years and/or the effects of inflation.

Focusing on built environment factors, there was a significant reduction in the average distance to the nearest transit stop, which decreased from 3.56 miles in 2006 to 0.9 miles in 2017, suggesting a substantial expansion of the transit network between these years. Regarding the distance to the CBD and the nearest activity center, the 2006 sample lived relatively farther from these locations, which is also reflected in its employment density. Although employment in the overall study area increased by 1.36 times from 1997 to 2006, the average TAZ-level employment density at the respondents' household locations decreased from 2.23 jobs per acre in 1997 to 1.8 jobs per acre in 2006. This decrease in employment density in 2006 compared to 1997 might be explained by households being located further from major employment centers. In 2017, both population and employment densities were higher than in the previous years. Lastly, the diversity index was relatively consistent across all three years.

**Table 1:** Variable description and sample statistics

| Variable Code | Variable Description | | Descriptive Statistics[1] | | |
| --- | --- | --- | --- | --- | --- |
| | | | 1997 | 2006 | 2017 |
| **Dependent Variable** | | | | | |
| VMT_Person | Individual daily vehicle miles traveled (VMT) | | 23.71 (33.9) | 20.28 (21.4) | 19.25 (19.4) |
| **Explanatory Variables** | | | | | |
| *Personal characteristics related variables* | | | | | |
| Sex | Sex of the respondent | Female* | 51.3% | 52.5% | 53.7% |
| | | Male | 48.7% | 47.5% | 46.3% |
| Ethnicity | Race or ethnicity of the respondent | Non-white/Caucasian* | 29.3% | 32.6% | 34.3% |
| | | White/Caucasian | 70.7% | 67.4% | 65.7% |
| Age | Age of the respondent in years | | 32.26 (20.5) | 37.56 (23.7) | 37.15 (22.8) |
| Disability | Whether the respondent has a transportation disability | No* | 95.1% | 95.5% | 95.5% |
| | | Yes | 4.9% | 4.5% | 4.5% |
| Licensed_Driver | Whether the respondent is a licensed driver | No* | 30.5% | 32.6% | 30.1% |
| | | Yes | 69.5% | 67.4% | 69.9% |
| Work_Hr_Wkly | Number of hours respondent worked weekly | | 20.20 (20.8) | 16.05 (20.7) | 18.24 (20.5) |
| Flex_Time | Whether the respondent's work hours are flexible | Unemployed* | 46.9% | 59.2% | 52.1% |
| | | Fixed | 28.4% | 24.0% | 27.9% |
| | | Flexible | 24.7% | 16.8% | 20.0% |

| Variable | Description | | | | |
|---|---|---|---|---|---|
| Home_Office_Wkly | Number of days respondent worked from home weekly | | 0.13 (0.7) | 0.27 (1.1) | 0.43 (1.3) |
| Student | Whether the respondent is a student | No* | 67.0% | 73.8% | 70.6% |
| | | Yes | 33.0% | 26.2% | 29.4% |
| *Household characteristics related variables* | | | | | |
| HH_Size | Number of persons living in the respondent's household | | 3.26 (1.5) | 3.53 (1.7) | 3.23 (1.4) |
| No_Employed | Number of employed persons living in the respondent's household | | 1.56 (0.9) | 1.36 (0.9) | 1.46 (0.9) |
| No_Vehicle | Number of motorized vehicles available in the respondent's household | | 1.96 (0.9) | 2.02 (0.9) | 2.05 (0.9) |
| No_Bike | Number of bicycles available in the respondent's household | | 1.41 (1.6) | 1.46 (1.7) | 1.49 (1.7) |
| HH_Selection | Factors influencing the respondent's current household location choice | Non-access factor* | 86.1% | 72% | 66.9% |
| | | Access factor | 13.9% | 28% | 33.1% |
| HH_Income | Yearly income of all members of the respondent's household ($ thousand) | | 46.27 (33.5) | 59.21 (42.0) | 81.76 (55.1) |
| *Built environment characteristics related variables* | | | | | |
| *Household level variables* | | | | | |
| Dist_transit | Distance from respondent's household location to the nearest transit stop in miles | | 3.61 (6.1) | 3.56 (5.7) | 0.89 (1.1) |
| Dist_CBD | Distance from the respondent's household location to the downtown (CBD) in miles | | 10.40 (8.5) | 11.89 (8.6) | 10.45 (6.9) |
| Dist_Actvy_Center | Distance from respondent's household location to the nearest activity center in miles | | 2.52 (2.3) | 3.15 (2.7) | 3.10 (2.2) |
| *TAZ level variables* | | | | | |
| PopDEN | Population density in the respondent's household TAZ (persons/acre) | | 5.46 (5.0) | 5.44 (4.4) | 6.41 (5.8) |
| EmplyDEN | Employment density in the respondent's household TAZ (jobs/acre) | | 2.23 (7.0) | 1.80 (4.1) | 2.85 (15.7) |
| Diversity[2] | 3-tier employment (basic, retail, and service) and household entropy | | 0.56 (0.2) | 0.58 (0.2) | 0.55 (0.2) |
| **Sample Size** | | | | | |
| Individual Sample | | | 4459 | 3020 | 6252 |
| Household Sample | | | 1854 | 1128 | 2459 |

\* These values of the categorical variables are designated as the reference category for modeling purposes.

[1] For categorical variables, percentages are provided. For numeric variables, the table shows means and standard deviations (in brackets).

[2] Diversity is calculated based on the following equation:

Diversity = [(HH/TA)*ln(HH/TA) + (BE/TA)*ln(BE/TA) + (RE/TA)*ln(RE/TA) + (SE/TA)* ln(SE/TA)] / ln(4)

Here: HH = Number of households in the TAZ, BE = Number of basic employment in the TAZ, RE = Number of retail employment in the TAZ, SE = Number of service employment in the TAZ, and TA = Total employment plus number of households.

# 3. Modeling approach

In this study, we aimed to integrate machine learning and inferential modeling techniques. More specifically, we first explored the non-linear relationships between built environment factors and individual daily VMT using a machine learning method, as this technique is highly effective for identifying non-linear relationships without presupposition. After examining the results of machine learning approach, we incorporated the identified non-linear relationships into inferential models. The advantage of inferential techniques is that they provide the magnitude and significance of the relationship, which cannot be directly obtained through a machine learning approach.

Before developing the models, we checked and addressed missing values and outliers in the three datasets. We also assessed multicollinearity among the explanatory variables using variance inflation factor (VIF) statistics. All the VIF values for the three datasets were found to be less than 5, indicating that multicollinearity was not an issue. In terms of modeling, we first conducted machine learning analyses and then, based on the findings from these models, performed inferential analyses, which we discuss below.

## 3.1 Machine Learning

In recent travel behavior studies—such as those investigating driving, mode choice, emissions, and transit ridership—Random Forest (RF) and Gradient Boosted Decision Trees (GBDT) are among the most widely used machine learning techniques, particularly for exploring non-linear relationships (Aghaabbasi & Chalermpong, 2023; Ashik et al., 2024; Cheng et al., 2019; Ding, Cao, & Næss, 2018; Hatami et al., 2023; Li & Kockelman, 2022). Both models are ensemble learning techniques that generate results by constructing multiple decision trees. RF is a bagging method that uses bootstrapping to resample the original data and trains multiple decision tree models on different subsets of the data, which helps reduce variance and prevent overfitting (Cheng et al., 2019). In contrast, GBDT is a boosting method that sequentially builds decision tree models, each focusing on correcting the errors of the previous model, thereby improving model accuracy and reducing bias (Ding, Cao, & Næss, 2018). Both techniques are highly efficient at handling large datasets and uncovering complex, non-linear relationships.

Additionally, we identified a recent algorithm called Extreme Gradient Boosting Model (XGBoost), an advanced variant of boosting technique, which is relatively new in travel behavior studies but may have the potential to perform as well as, or even better than, GBDT. Therefore, in our study, we utilized three machine learning algorithms—RF, GBDT, and XGBoost—to predict individual daily VMT, estimate the relative contribution of each explanatory factor, and explore the non-linear relationships between built environment factors and individual daily VMT.

To develop and assess the machine learning models, we first split our datasets into training and testing sets, with 20% of the data reserved for testing. A total of nine machine learning models were developed using the three algorithms (RF, GBDT, and XGBoost) across three separate datasets corresponding to the years 1997, 2006, and 2017. These models were implemented in Python using the scikit-learn library. To optimize the performance of these models, hyperparameters were tuned using GridSearchCV. The search ranges for hyperparameter optimization and the optimal hyperparameters identified are presented in **Table 2**. The tuning process was based on minimizing mean squared error (MSE) statistics, and a five-fold cross-validation (CV) procedure was applied.

After developing the models, we calculated the root mean squared error (RMSE) and $R^2$ statistics for each model to evaluate their performance (**Table 2**). A model is considered superior if it exhibits a lower RMSE and a higher $R^2$ compared to the others. Our results indicate that while the RF algorithm performs comparably to the GBDT algorithm, the GBDT slightly outperforms RF across all cases. Conversely, XGBoost did not perform as well as the other two algorithms. Consequently, we identified GBDT algorithm as the best-performing machine learning technique for this study. Among the GBDT models, the $R^2$ statistics were the highest for the models corresponding to the years 1997 and 2006 (33%), while the model for 2017 showed a slightly lower $R^2$ of 29% (**Table 2**). This suggests that the explanatory variables account for approximately 33% of the variance in individual daily VMT for the years 1997 and 2006, and 29% for the year 2017.

Given that the GBDT algorithm proved to be the most effective for our data, we used it to explore the relative importance of each explanatory variable in predicting individual daily VMT for all the three years (**Figure 2**). To investigate the non-linear relationships between the dependent variable and explanatory variables, Partial Dependence Plots (PDP) and Accumulated Local Effects (ALE) plots are commonly used in literature (Ashik et al., 2024; Ding, Cao, & Næss, 2018; Shao et al., 2023). Both methods have their own advantages and disadvantages: while PDPs are straightforward to compute and interpret, they do not perform well when multicollinearity is present among explanatory variables (Shao et al., 2023; W. Zhang et al., 2022b). On the other hand, ALE plots are more robust to multicollinearity but can be more challenging to interpret (Ashik et al., 2024; Shao et al., 2023). Since multicollinearity was not detected among our explanatory variables, we chose to explore the non-linear relationships between the explanatory variables—including built environment factors—and the dependent variable by generating PDPs based on the GBDT models. **Figures 3** and **Figure 4** illustrate the non-linear relationships between built environment factors and travel, while **Appendix A** presents the non-linear relationships involving other numeric explanatory variables.

**Table 2:** Hyperparameter tuning and performance of the models

| Model Type | Hyperparameter Optimization | | 1997 | 2006 | 2017 |
|---|---|---|---|---|---|
| | **Hyperparameter** | **Search Range** | | | |
| Random Forest | Number of trees | 100-1200 | 300 | 1000 | 1000 |
| | Maximum depth | 1-20 | 9 | 15 | 15 |
| | Maximum features | 'auto', 'sqrt', 'log2', None | 'sqrt' | 'sqrt' | 'sqrt' |
| GBDT | Number of estimators | 100-1200 | 550 | 1000 | 1000 |
| | Learning rate | 0.001-0.1 | 0.01 | 0.01 | 0.01 |
| | Maximum depth | 1-20 | 5 | 7 | 7 |
| | Maximum features | 'auto', 'sqrt', 'log2', None | 'sqrt' | 'sqrt' | 'sqrt' |
| | Subsample | 0.5-1.0 | 0.9 | 0.8 | 0.9 |
| XGBoost | Number of estimators | 100-1200 | 300 | 700 | 200 |
| | Learning rate | 0.001-0.1 | 0.02 | 0.075 | 0.05 |
| | Maximum depth | 1-20 | 4 | 3 | 6 |
| | Minimum child weight | 1-10 | 6 | 7 | 6 |
| | Subsample | 0.5-1.0 | 1.0 | 1.0 | 0.9 |
| | Column sample by tree | 0.5-1.0 | 0.9 | 0.7 | 0.9 |
| | Gamma | 0-0.5 | 0.0 | 0.0 | 0.1 |
| | | | | | |
| **Performance Evaluation** | | | | | |
| Random Forest | RMSE | | 28.58 | 17.12 | 17.53 |
| | $R^2$ | | 0.32 | 0.32 | 0.26 |
| GBDT | RMSE | | 28.40 | 17.02 | 17.17 |
| | $R^2$ | | 0.33 | 0.33 | 0.29 |
| XGBoost | RMSE | | 28.61 | 17.76 | 17.37 |
| | $R^2$ | | 0.32 | 0.27 | 0.27 |

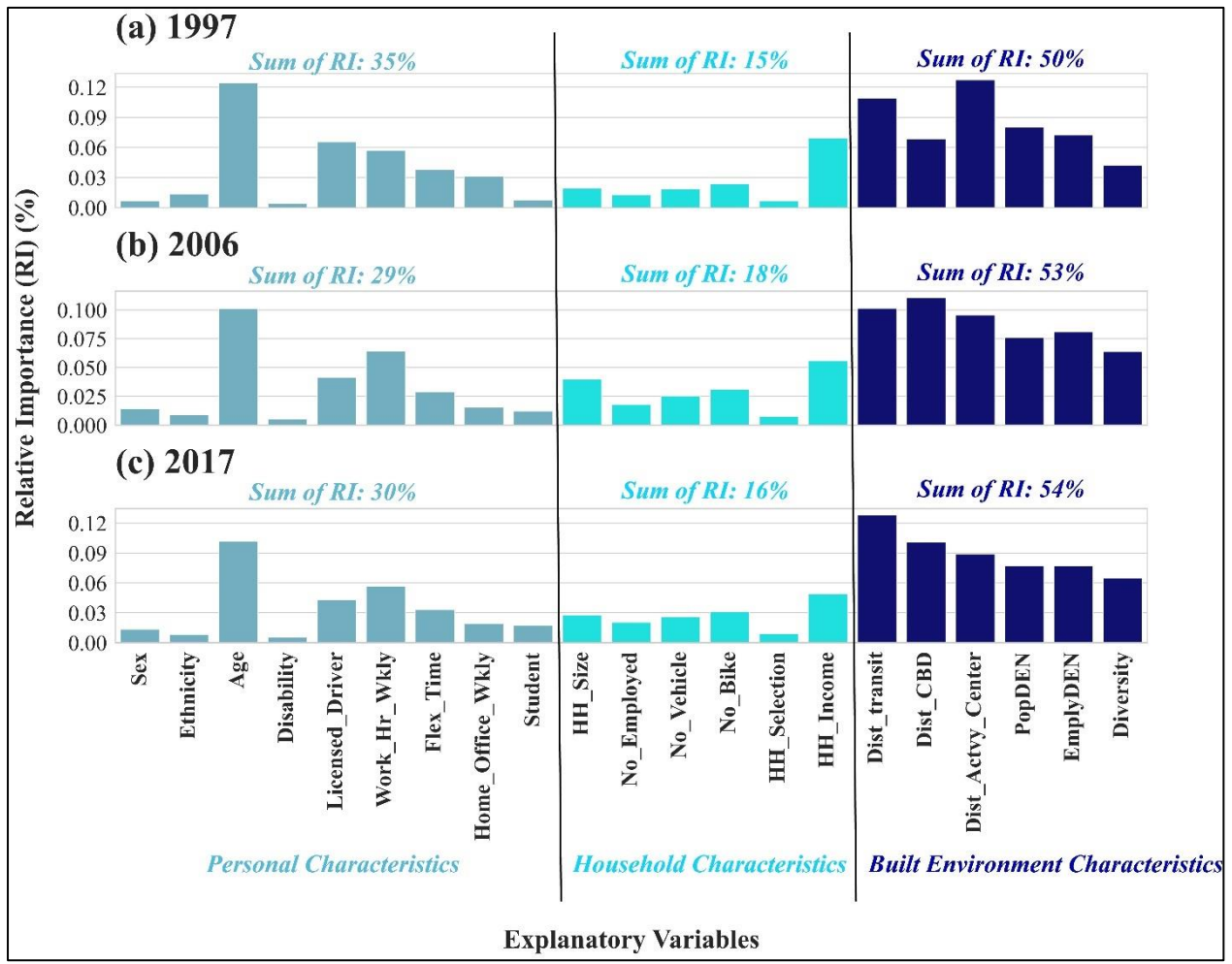

**Figure 2**: Relative importance of the explanatory variables to explain individual daily VMT (VMT_Person)

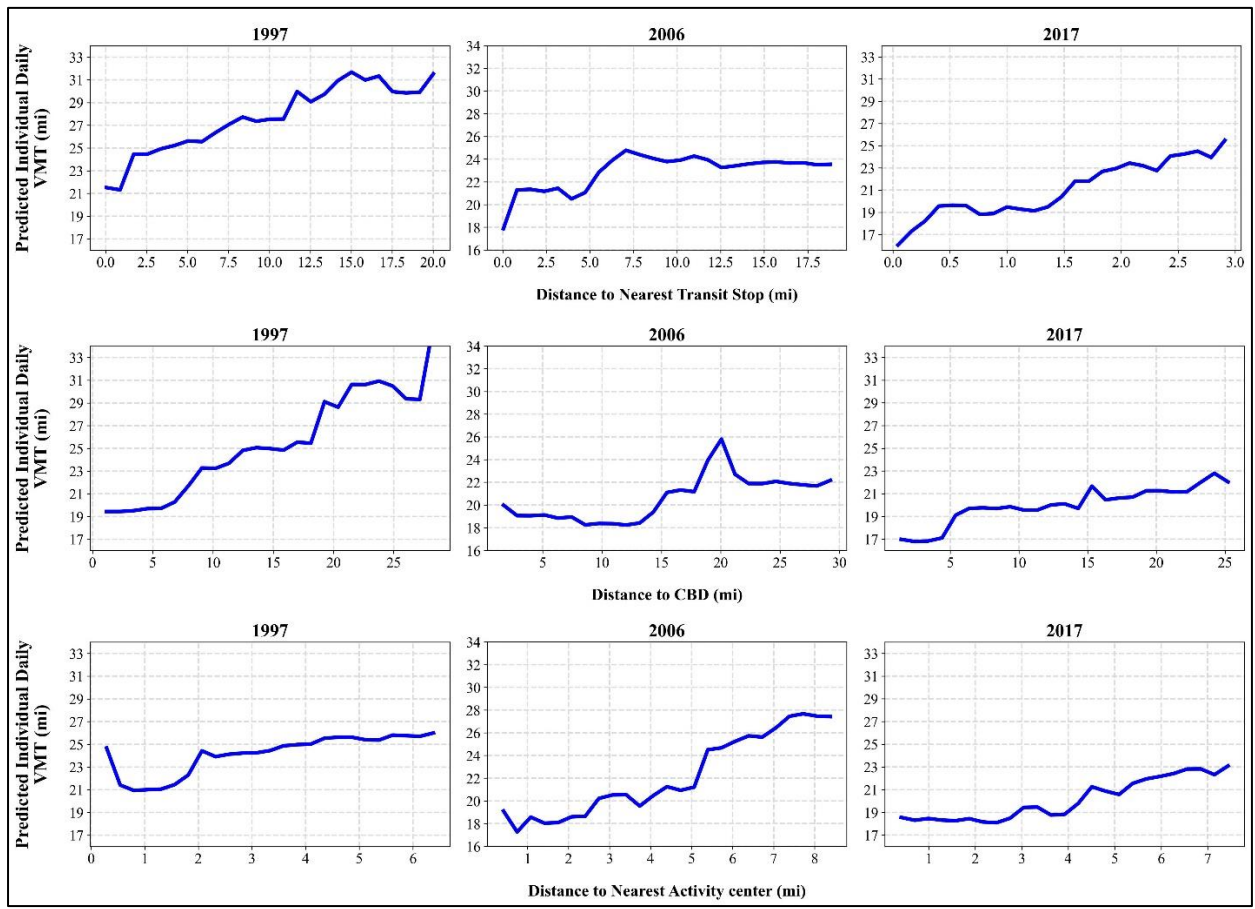

**Figure 3**: Non-linear effects of distance to nearest transit stop (Dist_transit), distance to CBD (Dist_CBD), and distance to nearest activity center (Dist_ Actvy_Center) on individual daily VMT (VMT_Person)

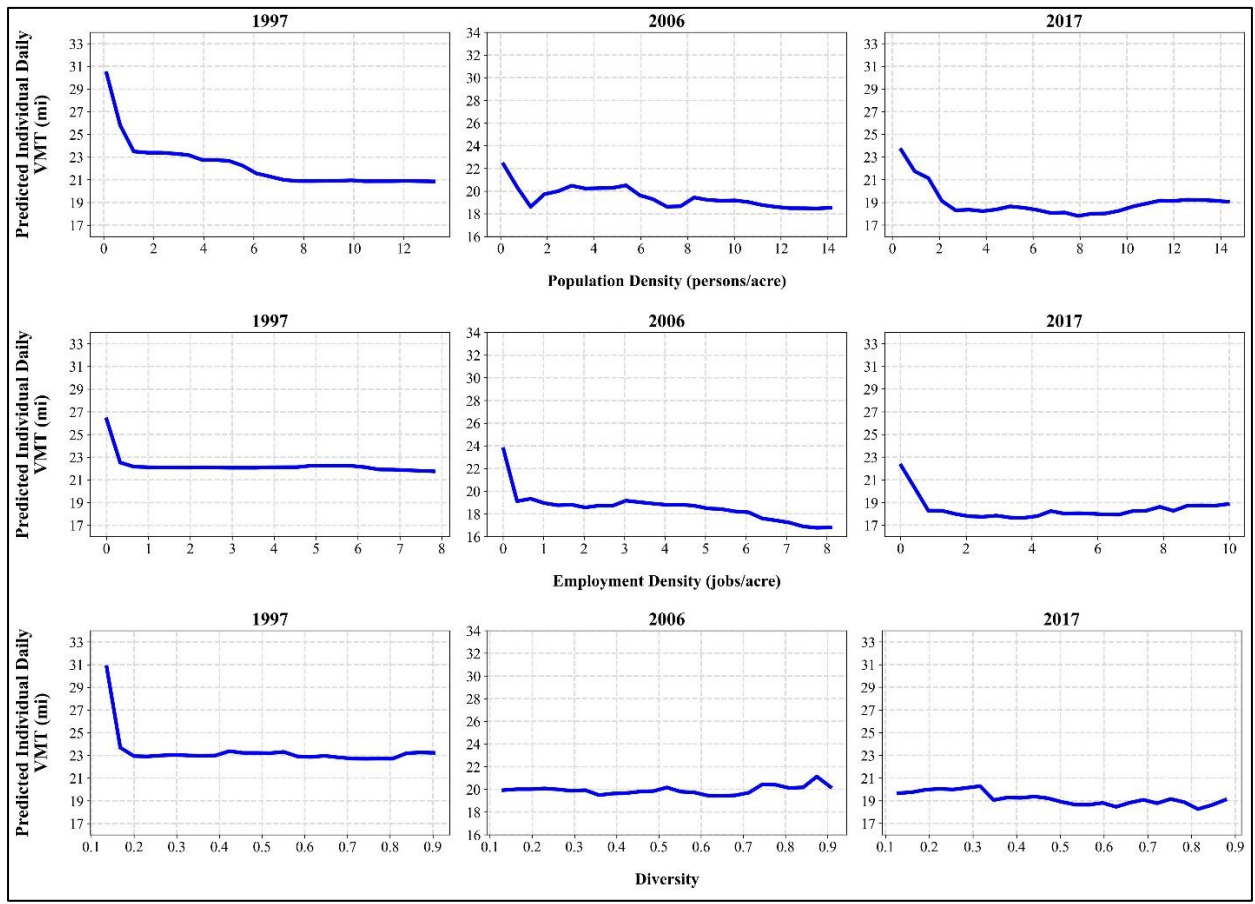

**Figure 4**: Non-linear effects of population density (PopDEN), employment density (EmplyDEN), diversity (Diversity) on individual daily VMT (VMT_Person)

## 3.2 Inferential Modeling

Since our dependent variable is continuous, a linear regression model might initially seem appropriate for our study. However, upon examining the structure of our data, we found it to be hierarchical: individuals are nested within households, and households are nested within TAZs. Due to this clear hierarchical structure, a single-level linear regression model is not ideal, as it assumes independence among observations—an assumption that does not hold in our case. As a result, using a single-level linear regression would lead to inefficient estimates for the study (Sabouri et al., 2020).

In such cases, hierarchical linear regression, also known as multilevel modeling, is the preferred approach. This method overcomes the limitations of traditional linear regression by accounting for the nested structure of the data, allowing for more accurate estimation of coefficients and standard errors (Bryk & Raudenbush, 1992; Sabouri et al., 2020). Consequently, we employed a multilevel modeling approach in our study. Specifically, we allowed only the intercept to vary and keep the slopes constant. Thus, we used a random intercept multilevel modeling technique to simplify model development and interpretation.

For each year's data, we developed a three-level random intercept model. The first level captures the effects of personal characteristics, the second level addresses household characteristics, and the third level incorporates TAZ characteristics. It is important to note that built environment variables such as Dist_transit, Dist_CBD, and Dist_Actvy_Center are associated with individual households and were included at the second level of the model for the years 2006 and 2017. However, for the 1997 data, these variables were linked to TAZs because all households within the same TAZ shared the same values; this was due to household locations being approximated by the TAZ centroid, as precise household coordinates were not available. Therefore, these variables were included at the third level for the 1997 model. Other built environment variables, such as PopDEN, EmplyDEN, and Diversity, were TAZ-level characteristics and were included at the third level across all three models.

Non-linear effects for the continuous variables were incorporated into the multilevel models using a piecewise regression method (Buscot et al., 2017; Toms & Lesperance, 2003). The knot(s) (breakpoint) for developing piecewise linear segments of a numeric variable were determined based on the findings of PDPs. When selecting the knot(s) for an explanatory variable, we examined the PDPs that depict the relationship between individual daily VMT and the corresponding explanatory variable across all three years. A value of the explanatory variable was identified as a knot if the relationship before and after that point showed visible differences. These knots were then generalized for each explanatory variable based on the results from all three years to avoid issues of overfitting or underfitting. Finally, we constructed piecewise linear segments for the explanatory variables based on these knots.

Take Dist_CBD as an example. First, we estimated the non-linear relationship between Dist_CBD and VMT_Person using PDPs (**Figure 3**). Upon examining this relationship, we observed that the nature and magnitude of the relationship before and after approximately 5 miles from the CBD differed in the years 1997 and 2017. Additionally, a distinct pattern emerged before and after approximately 15 miles from the CBD, particularly for the year 2006. There might also be another potential breakpoint at 20 miles from the CBD for 2006; however, including this could lead to an overfitting issue. Thus, we identified two knots for this variable: 5 and 15 miles. Using these two knots, we modeled the relationship between Dist_CBD and VMT_Person in three segments within the multilevel model: the first segment reflects the relationship within the 0 to 5 miles range, the second represents the 5 to 15 miles range, and the third represents distances beyond 15 miles.

The models were developed following a specific step by step procedure. In the first step, we developed a null model, and subsequently, we added variables related to individual and household characteristics, including their non-linear effects. Later, we incorporated built environment variables without accounting for non-linear effects. Finally, we included non-linear effects of the built environment factors into the model. At each step, we removed variables that were not statistically significant.

Model fit statistics for each step of the modeling process are presented in **Appendix B**. All fit statistics (AIC, BIC, log-likelihood, marginal $R^2$, and conditional $R^2$) indicate that the final models for all three years (developed in step 4) are significantly better than the models from the earlier steps. The standard deviations of the random effects at levels 2 and 3 significantly decreased in the models developed in the third step compared to those in the second step. This suggests that incorporating the effects of built environment factors at levels 2 and 3 helps explain a significant portion of the intercept variance at these levels. Furthermore, the standard deviations of the random effects decreased even further in the final models compared to the third-step models, indicating that including non-linear effects of built environment factors can further explain a significant portion of the intercept variance at levels 2 and 3. Overall, these findings demonstrate that models incorporating non-linear effects of built environment factors are superior, emphasizing the importance of including non-linear effects to improve model performance. The results of the developed models are presented in **Table 3**.

Here is an example of how to interpret the non-linear results. Consider the Dist_CBD variable for the year 1997 (**Table 3**). Within the 0 to 5 miles range from the CBD, an increase of 1 mile in distance decreases individual daily VMT by 0.69 miles. However, this effect is not statistically significant, indicating that within this range, changes in distance from the CBD do not significantly impact VMT. In contrast, within the 5 to 15 miles range, each additional mile significantly increases VMT by 0.85 miles. Beyond 15 miles from the CBD, each additional mile significantly increases VMT by 1.28 miles.

After developing the model, we estimated the elasticity of the built environment factors (**Table 4**). Elasticity is a unit-less measure that indicates the impact of a one-percent change in the explanatory variable on the percentage change of the dependent variable. For our model, elasticity is calculated as the regression coefficient multiplied by the mean value of the explanatory variable, divided by the mean value of the dependent variable (Ewing & Cervero, 2010).

**Table 3:** Result of multilevel models

| Explanatory Variables | 1997 | 2006 | 2017 |
|---|---|---|---|
| | β (SE) | β (SE) | β (SE) |
| **Fixed Effects** | | | |
| Intercept | -11.04 (4.88)** | 26.23 (5.27)*** | 16.33 (3.45)*** |
| Ethnicity [White/Caucasian] | 2.26 (1.29)* | | |
| Age [0-6] | 3.01 (0.51)*** | -2.13 (0.37)*** | -1.61 (0.26)*** |
| Age [6-20] | -0.18 (0.19) | 0.34 (0.14)*** | 0.05 (0.10) |
| Age [20-50] | 0.14 (0.06)** | 0.16 (0.05)*** | 0.19 (0.03)*** |
| Age [>50] | -0.18 (0.09)** | -0.32 (0.06)*** | -0.24 (0.04)*** |
| Disability [Yes] | | -3.31 (1.60)** | -2.27 (1.11)** |
| Licensed_Driver [Yes] | 16.10 (1.81)*** | 6.94 (1.28)*** | 8.33 (0.94)*** |
| Work_Hr_Wkly | | | 0.06 (0.03)** |
| Flex_Time [Fixed] | 6.88 (1.32)*** | 8.85 (1.02)*** | 3.26 (1.32)*** |
| Flex_Time [Flexible] | 7.44 (1.32)*** | 6.69 (1.16)*** | 4.76 (1.23)*** |
| Home_Office_Wkly [0-3] | | 0.23 (0.69) | 0.55 (0.42) |
| Home_Office_Wkly [>3] | | -3.22 (0.98)*** | -1.93 (0.56)*** |
| HH_Size [0-6] | | -0.80 (0.43)* | |
| HH_Size [>6] | | 4.88 (1.58)*** | |
| No_Employed | | -2.67 (0.69)*** | -1.82 (0.44)*** |
| No_Vehicle [0-2] | | | 1.95 (0.68)*** |
| No_Vehicle [2-4] | | | -1.40 (0.58)** |
| No_Vehicle [>4] | | | 3.32 (2.44) |
| No_Bike [0-6] | | 1.10 (0.36)*** | 0.52 (0.21)*** |
| No_Bike [>6] | | -2.61 (3.93) | -2.26 (1.19)* |
| HH_Selection [Access factor] | -3.53 (1.70)** | | -1.64 (0.61)*** |
| HH_Income [0-20] | -0.04 (0.18) | -0.01 (0.15) | -0.10 (0.14) |
| HH_Income [20-50] | 0.25 (0.07)*** | 0.10 (0.05)* | 0.10 (0.04)*** |
| HH_Income [>50] | 0.02 (0.03) | 0.04 (0.02)*** | -0.02 (0.01)** |
| Dist_transit [0-7] | | 1.50 (0.27)*** | 1.66 (0.38)*** |
| Dist_transit [>7] | | -0.40 (0.18)** | -5.67 (13.95) |
| Dist_ Actvy_Center [0-2] | 3.94 (1.62) ** | 1.29 (0.92) | |
| Dist_ Actvy_Center [>2] | -0.18 (0.49) | 1.06 (0.26) *** | |
| Dist_CBD [0-5] | -0.69 (0.79) | | 1.26 (0.31) *** |
| Dist_CBD [5-15] | 0.85 (0.27) *** | | 0.33 (0.11) *** |
| Dist_CBD [>15] | 1.28 (0.20) *** | | -0.27 (0.11) *** |
| PopDEN [0-2] | -6.60 (1.37) *** | -1.14 (1.00) | -2.90 (0.73) *** |
| PopDEN [>2] | -0.03 (0.19) | -0.22 (0.14) * | -0.01 (0.06) |
| EmplyDEN [0-1] | | | -3.76 (0.96) *** |
| EmplyDEN [>1] | | | 0.01 (0.02) |
| Diversity [0-0.2] | | -39.81 (18.66) ** | |
| Diversity [>0.2] | | -1.09 (2.37) | |
| **Random Effects (Standard Deviation)** | | | |
| Individual (Residual) | 23.1 | 14.8 | 14.5 |
| Household (Level 2) | 17.7 | 11.2 | 9.8 |
| TAZ (Level 3) | 7.1 | 0.01 | 0.6 |
| **Sample Size** | | | |
| Observation | 4459 | 3020 | 6252 |
| Household | 1854 | 1128 | 2459 |
| TAZ | 485 | 477 | 778 |

***Significant at 1% level, **Significant at 5% level, and * significant at 10% level

**Table 4**: Elasticity of Built Environment Factors in 1997, 2006, and 2017

| Built Environment Factors | 1997 | 2006 | 2017 |
|---|---|---|---|
| Dist_transit [0-7] | | 0.26*** | 0.08*** |
| Dist_transit [>7] | | -0.07** | -0.26 |
| Dist_ Actvy_Center [0-2] | 0.42** | 0.20 | |
| Dist_ Actvy_Center [>2] | -0.02 | 0.16*** | |
| Dist_CBD [0-5] | -0.30 | | 0.68*** |
| Dist_CBD [5-15] | 0.37*** | | 0.18*** |
| Dist_CBD [>15] | 0.56*** | | -0.15*** |
| PopDEN [0-2] | -1.52*** | -0.31 | -0.96*** |
| PopDEN [>2] | -0.01 | -0.06* | 0.00 |
| EmplyDEN [0-1] | | | -0.56*** |
| EmplyDEN [>1] | | | 0.00 |
| Diversity [0-0.2] | | -1.14** | |
| Diversity [>0.2] | | -0.03 | |

***Significant at 1% level, **Significant at 5% level, and * significant at 10% level

# 4. Results and Discussion

## 4.1 Can the built environment sustain as a tool for controlling travel?

Based on the GBDT models, the relative importance of explanatory variables in predicting individual daily VMT was analyzed by estimating feature importance (**Figure 2**). The analysis revealed that built environment characteristics combinedly contributed 50%, 53%, and 54% to the predictive performance of the models in 1997, 2006, and 2017, respectively. This indicates that built environment factors exert a greater influence on the predictive capabilities of the GBDT models than the combined effects of personal and household characteristics. Specifically, the relative importance of individual characteristics was 35%, 29%, and 30% in 1997, 2006, and 2017, respectively, which is consistently higher than the relative importance of household characteristics, which stood at 15%, 18%, and 16% across the same years. This trend demonstrates that built environment characteristics are consistently the most influential and stable predictors of individual daily VMT across all years studied.

Within the built environment characteristics, the distance to the nearest transit stop, the distance to CBD, and the distance to the nearest activity center were the most influential factors in all three years, except for 1997, where the distance to CBD had a moderate effect compared to the other two factors. Population density and employment density also had a moderate impact on VMT in all three years, whereas the diversity index consistently emerged as the least significant factor, particularly in 1997.

In summary, although the relative importance of specific built environment factors fluctuated over the years, the overall influence of built environment characteristics on individual daily VMT remained consistently high and increased marginally over time. These findings underscore the built environment's role as a robust and sustainable tool for mobility management in the long term.

## 4.2 How have the built environment and travel relationship and its magnitude evolved over time?

### 4.2.1 Distance to nearest transit stop (Dist_transit)

The PDPs reveal that the relationship between the distance to the nearest transit stop and individual daily VMT was positive and nearly linear for the years 1997 and 2017, indicating that as the distance from a transit stop increases, VMT also increases (**Figure 3**). However, in 2006, this relationship was positive and moderately linear only up to a 7-mile threshold, beyond which the distance to the nearest transit stop had no mentionable influence on VMT.

The multilevel models indicate that the distance to the nearest transit stop was not a significant predictor of VMT in 1997 but became significant in both 2006 and 2017 (**Table 3**). Specifically, within the 0 to 7-mile range, greater distances to transit stops were significantly associated with higher VMT in both 2006 and 2017. In this range, an increase of one mile in the distance from the nearest transit stop to a household location led to an increase in individual daily VMT by 1.5 miles in 2006 and 1.66 miles in 2017. However, beyond the 7-mile threshold, the relationship became negative, suggesting that longer distances from transit stops correspond to lower VMT. The impact magnitude was relatively small beyond 7 miles in 2006 and was insignificant in 2017.

Within the 7-mile threshold, the magnitude of impact was greater in 2006 compared to 2017, as indicated by the elasticity statistics. An elasticity of 0.26 in 2006 and 0.08 in 2017 suggests that a 100% decrease in the distance to the nearest transit stop would lead to a 26% and 8% decrease in individual daily VMT, respectively, within this range (**Table 4**). The reduced impact magnitude in 2017, compared to 2006, could be attributed to the extensive expansion of the transit network, which may have lessened the influence of this factor on controlling VMT in more recent years (see **Table 1**: the mean Dist_transit was 3.56 miles in 2006 and 0.89 miles in 2017). However, this distance is still well above the recommended 0.25- and 0.5-mile threshold, indicating room for further improvement.

Most previous studies have found a similar relationship between distance to transit and VMT, which is consistent with our findings (Cao et al., 2007; van de Coevering et al., 2021). Through meta-analyses of existing literature, Ewing & Cervero (2010) and Stevens (2017) identified a relatively small elasticity of 0.05 for this variable. However, our study found a very higher impact magnitude for Austin in 2006 (0.26) and slightly high magnitude in 2017 (0.08) within 7-mile threshold, which suggests the need to explore threshold effects to estimate the accurate magnitude of impact, the possibility that access to transit has a greater influence in the study area, or a combination of both factors.

In summary, within the 7-mile threshold, the distance to transit is a highly influential factor, particularly in 2006 and 2017, where lower accessibility to transit significantly increases VMT. Additionally, our study uniquely identifies that beyond this threshold, the distance to transit in Austin ceases to be an effective tool for controlling VMT, indicating the presence of a threshold effect.

### 4.2.2 Distance to CBD (Dist_CBD)

Within the 0 to 5-mile range from the CBD, individual daily VMT showed minimal variation with changes in distance to the CBD for 1997 and 2006, as indicated by the PDPs (**Figure 3**). However, for 2017, a slight increase in VMT was observed within this range with the distance from the CBD increased. The results of the multilevel models align with these findings (**Table 3**). The effect of distance to the CBD was not significant for 1997 and 2006, suggesting that within this range, changes in distance had no substantial impact on VMT. However, in 2017, an increase of 1 mile in distance from the CBD to a household significantly increased individual daily VMT by 1.26 miles. The elasticity for this variable suggests that doubling the distance to the CBD (a 100% increase) would result in a 68% increase in daily VMT, demonstrating a significant impact magnitude (**Table 4**).

As indicated by the PDPs, in the 5 to 15-mile range, VMT increased with greater distance from the CBD in 1997 (**Figure 3**). However, no clear relationship was found between VMT and distance to the CBD for 2006 and 2017 within this range. The multilevel models produce results that closely align with the PDPs (**Table 3**). Specifically, an increase of 1 mile in distance from the CBD significantly increased VMT by 0.85 miles in 1997 and by 0.33 miles in 2017. Additionally, the effect was not significant for 2006, indicating that within this range, the distance from the CBD did not influence individual daily VMT in that year. The elasticity values of 0.37 in 1997 and 0.18 in 2017 suggest that doubling the distance within this range would increase individual daily VMT by 37% and 18%, respectively (**Table 4**). These statistics also indicate that the magnitude of impact was twice as high in 1997 compared to 2017.

According to the PDPs, beyond 15 miles from the CBD, a relatively positive linear relationship was found between VMT and distance to the CBD in 1997 (**Figure 3**). However, for 2006 and 2017, this relationship varied slightly without a clear directional trend. The multilevel models also reflect these findings (**Table 3**). Beyond 15 miles from the CBD, an increase of 1 mile in distance significantly increased daily VMT by 1.28 miles in 1997. In contrast, this variable was not significant in 2006. In 2017, although the variable was significant, it had a negative effect on VMT, indicating that for every 1 mile increase in distance from the CBD beyond 15 miles, VMT decreased by 0.27 miles. The elasticity statistics of 0.56 in 1997 and -0.15 in 2017 suggest that doubling the distance within this range would increase VMT by 56% in 1997 and decrease it by 15% in 2017 (**Table 4**).

From the above findings, we can see that the distance from the CBD was one of the most significant factors in controlling VMT in both 1997 and 2017. The results predominantly indicate that the farther from the CBD (signifying lower regional accessibility), the higher the VMT tended to be. This underscores the importance of providing housing opportunities closer to the CBD rather than in peripheral areas to reduce travel. Previous studies have also identified distance from the CBD as a critical factor influencing travel behavior. Meta-analyses by Ewing & Cervero (2010) and Stevens (2017) reported the highest elasticity statistics for this variable in controlling VMT, with values of 0.22 and 0.63, respectively, showing strong consistency with our results.

Interestingly, beyond 15 miles from the CBD, VMT decreased as the distance increased in 2017. This decrease could be attributed to factors such as higher rates of telecommuting and online shopping among residents living more than 15 miles from the CBD in 2017 compared to earlier years, reducing their need to travel. This trend is also evident in our study results for the variable "Home_Office_Wkly [>3]," which shows that individuals who work from home more frequently (three or more days per week) tend to have lower VMT. This variable was not significant in 1997. Previous studies have similarly found that people living far from the CBD are more likely to telecommute and engage in online shopping (Zhu et al., 2023). In summary, residents living farther from the CBD are more likely to engage in virtual activities, consequently reducing their travel needs.

### 4.2.3 Distance to the nearest activity center (Dist_ Actvy_Center)

From the PDPs, we observe that within the 0 to 2-mile range from the nearest activity center, VMT increases sharply in 1997 (**Figure 3**). Beyond this threshold, the increase in VMT continues but at a slower, linear rate. In contrast, for 2006 and 2017, within the 2-mile range, distance to the nearest activity center does not appear to significantly influence VMT. However, beyond the 2-mile threshold, a linear positive relationship between VMT and the distance to the nearest activity center becomes evident, particularly for 2006.

The multilevel models confirm these findings, showing that distance to the nearest activity center was significant in 1997 and 2006, with its impact diminishing after 2006. Specifically, within the 2-mile range from the nearest activity center, longer distances are associated with higher VMT in 1997, with a notable increase of 3.94 miles in VMT for every additional mile from the activity center. Higher elasticity statistics indicate that a 100% increase in the distance to the nearest activity center results in a 42% increase in individual daily VMT within the 2-mile threshold. In contrast, in 2006 and 2017, distance to the nearest activity center does not significantly affect VMT within this range. Beyond the 2-mile threshold, longer distances to activity centers positively influence VMT, but significant effects are observed only in 2006. The elasticity beyond the 2-mile threshold was 0.16 for 2006, indicating that a 100% decrease in the distance to the nearest activity center would reduce VMT by 16%.

In sum, distance to the nearest activity center had a significant influence on individual daily VMT and its impact diminished in 2017. Individuals tend to travel more when living farther from activity centers (low local accessibility). This result is consistent with our expectations and aligns with previous studies (Cao et al., 2007; M. Zhang & Zhang, 2020).

### 4.2.4 Population density (PopDEN)

From the PDPs, we observe that population density exhibits a similar kind of relationship with individual daily VMT across all three years (**Figure 4**). VMT decreases significantly with increasing population density up to a threshold of 2 persons per acre, indicating that denser areas up to this point encourage less travel. Beyond this threshold, the impact of population density on VMT weakens.

The multilevel models confirm that population density is statistically significant for all three years, showing a negative relationship with VMT (**Table 3**). Specifically, within the 0 to 2 persons per acre range, increasing population density significantly reduces VMT. For every additional unit increase in population density within this range, VMT decreases by 6.6 miles in 1997, 1.14 miles in 2006, and 2.9 miles in 2017. However, beyond the 2 persons per acre threshold, the effect on VMT is significant only in 2006, with a decrease of 0.03 miles in 1997, 0.23 miles in 2006, and 0.01 miles in 2017 for each additional unit increase in population density.

Elasticity values of -1.52 in 1997, -0.31 in 2006, and -0.96 in 2017 suggest that doubling the population density would result in a 152%, 31%, and 96% decrease in individual daily VMT, respectively, within the 2 persons per acre threshold (**Table 4**). Beyond this threshold, the elasticity values indicate a 1%, 6%, and 0% reduction in VMT in 1997, 2006, and 2017, respectively, due to a doubling of population density, which is less influential compared to the impact within the 2 persons per acre threshold.

Population density exerts a negative influence on VMT across all three years, particularly within the 0 to 2 persons per acre threshold. This result aligns with expectations and is consistent with previous research suggesting that higher population density can reduce VMT. However, meta-analyses by Ewing & Cervero (2010) and Stevens (2017) reported relatively low elasticity values for this variable in controlling VMT, with reported values of 0.04 and 0.22, respectively. These values indicate a lower magnitude of impact compared to our findings within the 2 persons per acre threshold. Furthermore, a study by (M. Zhang & Zhang, 2020) reported an elasticity of 0.026 for population density in Austin, which is notably lower than our findings for the 2 persons per acre threshold. This discrepancy may arise from the lack of consideration of threshold effects in previous studies, which potentially undermined the policy relevance of population density. Our study, however, clearly demonstrates that within the 2 persons per acre threshold, population density significantly impacts VMT and, therefore, has considerable policy relevance. To provide additional context and emphasize this point, it is important to note that, in 2017, 51% of the TAZs had a population density below this threshold.

### 4.2.5 Employment Density (EmplyDEN)

From the PDPs, we can observe that individual daily VMT decreases with an increase in employment density up to a threshold of 1 job per acre in all three years (**Figure 4**). Beyond this threshold, employment density appears to have no influence on VMT in 1997 and 2017, while a slightly negative relationship is observed in 2006.

When compiling results from the multilevel models, the effect of employment density is not statistically significant in 1997 and 2006 (**Table 3**). However, it becomes statistically significant in 2017, suggesting that its influence may have emerged over time. Within the 1 job per acre threshold, a one-unit increase in employment density decreases individual daily VMT by 3.76 miles, indicating that living in areas with more job opportunities nearby reduces travel. The elasticity statistic is -0.56, meaning that doubling the number of jobs within this threshold decreases VMT by 56% (**Table 4**). Beyond this range, however, employment density does not have any significant influence.

Although previous studies suggests higher employment density is associated with lower VMT like our study, Ewing & Cervero (2010) and Stevens (2017) reported low elasticity values for employment density in relation to VMT reduction, with values of 0.0 and 0.07, respectively. These figures suggest a smaller impact compared to our findings within the 1 job per acre threshold in 2017. This difference could be due to the previous studies not accounting for threshold effects. Our research highlights that employment density within the 1 job per acre threshold is highly relevant for policy considerations. To further contextualize this, it is noteworthy that in 2017, 55% of the TAZs had a job density below this threshold.

### 4.2.6 Diversity

From the PDPs, it is evident that diversity generally does not influence individual daily VMT, except within the 0 to 0.2 entropy index range for the year 1997 (**Figure 4**). Within this threshold, increasing diversity substantially decreases VMT.

The multilevel models show that diversity was not significant in 1997 and 2017 but was significant in 2006, suggesting that its impact emerged in 2006 and then diminished afterward (**Table 3**). Specifically, within the 0.2 diversity threshold in 2006, a 0.1 unit increase in diversity entropy reduces VMT by 3.98 miles. The elasticity value of -1.14 indicates that doubling diversity within this range could reduce VMT by 114%, which is substantial (**Table 4)**. Beyond this threshold, diversity does not have a significant effect on VMT.

In summary, diversity appears to be a less influential factor in the study area. However, within the 0 to 0.2 threshold, higher diversity can reduce VMT, indicating that mixed-use areas promote shorter travel distances and fewer trips. When comparing these results with previous ones, while Ewing & Cervero (2010) and Stevens (2017) reported very small and counterproductive elasticity values, M. Zhang & Zhang (2020) estimated an elasticity value of 0.55 for Austin, which might be comparable with our study. Although this value is less compared to our findings within the 1 job per acre threshold, it may be due to the lack of consideration of threshold effects, as observed with other variables.

## 5. Summary and Conclusion

This study examines the dynamic relationship between the built environment and travel behavior, specifically individual daily VMT, over a 20-year period in Austin, Texas. Utilizing three waves of household travel survey data from 1997, 2006, and 2017, it integrates the strengths of machine learning and inferential modeling to introduce a novel approach for accurately and efficiently understanding the non-linear and threshold effects of the built environment on travel behavior.

Through a repeated cross-sectional approach, this study reveals that built environment factors consistently account for 50% or more of the relative importance in predicting individual daily VMT, exerting a greater influence than personal and household characteristics combined across all three years. This finding underscores the built environment's long term potential as a tool for managing travel. Notably, it provides robust evidence of the enduring effectiveness of built environment factors on travel behavior—an assumption often made by researchers but rarely supported by empirical evidence.

The study further contributes to the field by identifying key non-linear relationships and threshold effects that are critical for policymaking. Among the built environment factors, distance to the nearest transit stop emerged as a dominant predictor of VMT, especially in 2006 and 2017. For distances within a 7-mile threshold, increased transit accessibility was associated with reduced daily VMT; beyond this threshold, transit accessibility had no significant effect. However, due to Austin's extensive transit network expansion after 2005, the influence of this variable was less pronounced in 2017 compared to 2006.

Regional and local destination accessibility were also critical. Both the distance to the CBD (regional accessibility) and the nearest activity centers (local accessibility) were more influential in 1997 than in later years. In 2006, local accessibility became more important, while regional accessibility gained significance in 2017. Lower regional accessibility generally increased VMT, except in 2017, when increasing distance beyond the 15-mile threshold was associated with lower VMT. Local accessibility had a significant impact within a 0 to 2-mile range in 1997 and 2006, indicating that greater local accessibility reduces VMT. The influence of this factor diminishes beyond the threshold.

Population density consistently impacted VMT across all years, with a notable reduction in VMT observed up to a threshold of 2 persons per acre. However, as density nearly doubled over the study period, the magnitude of its impact diminished to some extent in 2017 compared to 1997. Similarly, employment density, particularly in 2017, was influential, reducing VMT within a threshold of 1 job per acre. Diversity, while less influential overall, significantly reduced VMT up to an entropy index of 0.2, particularly in 2006.

The results confirm prior hypotheses while providing more nuanced evidence regarding the magnitude and nature of the relationships between the built environment and travel behavior through the identification of non-linear effects and thresholds. Policy implications include promoting smart growth strategies such as enhancing transit accessibility in areas where the average distance to transit falls within the 7-mile threshold, encouraging population and employment densification up to 2 persons per acre and 1 job per acre, respectively, and discouraging single-use developments by achieving a mixed-use entropy index of 0.2. Additionally, avoiding sprawl by focusing development within 5 miles of the CBD and 2 miles of local activity centers is critical. As dynamic results show that transit accessibility and population density approach their optimal impact levels, greater emphasis should be placed on improving employment density and land-use diversity. These factors hold untapped potential and could become key drivers in reducing car dependency in Austin in the coming decades.

These results are likely generalizable to cities in developed countries with characteristics similar to Austin. The study concludes by advocating for additional research in varied contextual settings using the proposed integrated approach. This will not only improve the generalizability of findings but also advance the understanding of non-linear dynamics in the built environment-travel relationship, providing a stronger foundation for evidence-based policymaking.

**Funding source**

This study is funded by the U.S. Department of Transportation via the Center for Understanding Future Travel Behavior and Demand (TBD) under Grant Numbers 69A3552344815 and 69A3552348320.

**CRediT authorship contribution statement**

**Niaz Mahmud Zafri**: conceptualization, data curation, formal analysis, investigation, methodology, software, writing – original draft, and writing – review and editing; **Ming Zhang**: conceptualization, data curation, funding acquisition, methodology, project administration, supervision, writing – review and editing

**Declaration of competing interest**

The author(s) declared no potential conflicts of interest with respect to the research, authorship, and/or publication of this article.

# Appendix A

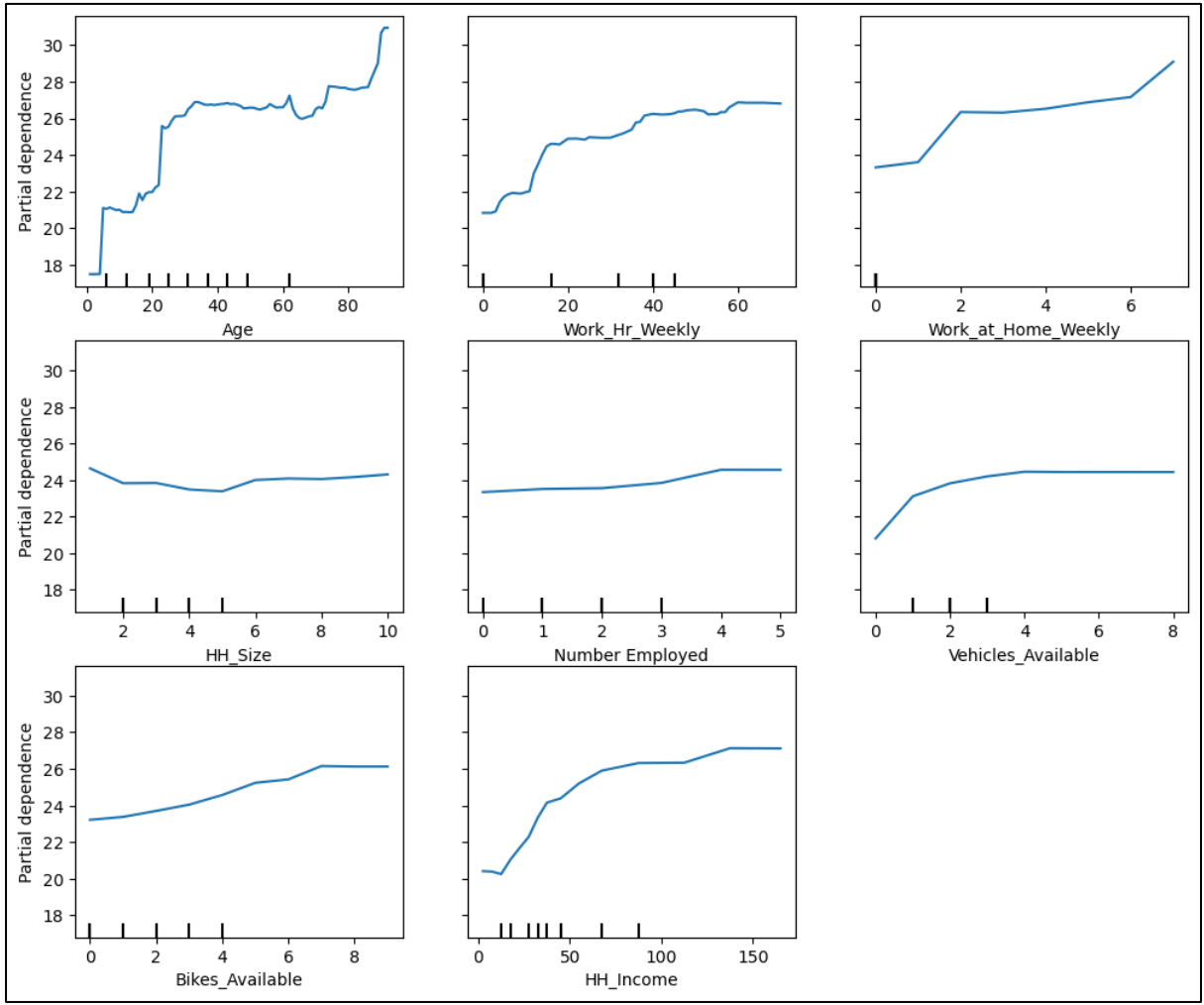

**Figure A1:** Relationship between VMT_Person and non-built environment continuous explanatory variables in 1997.

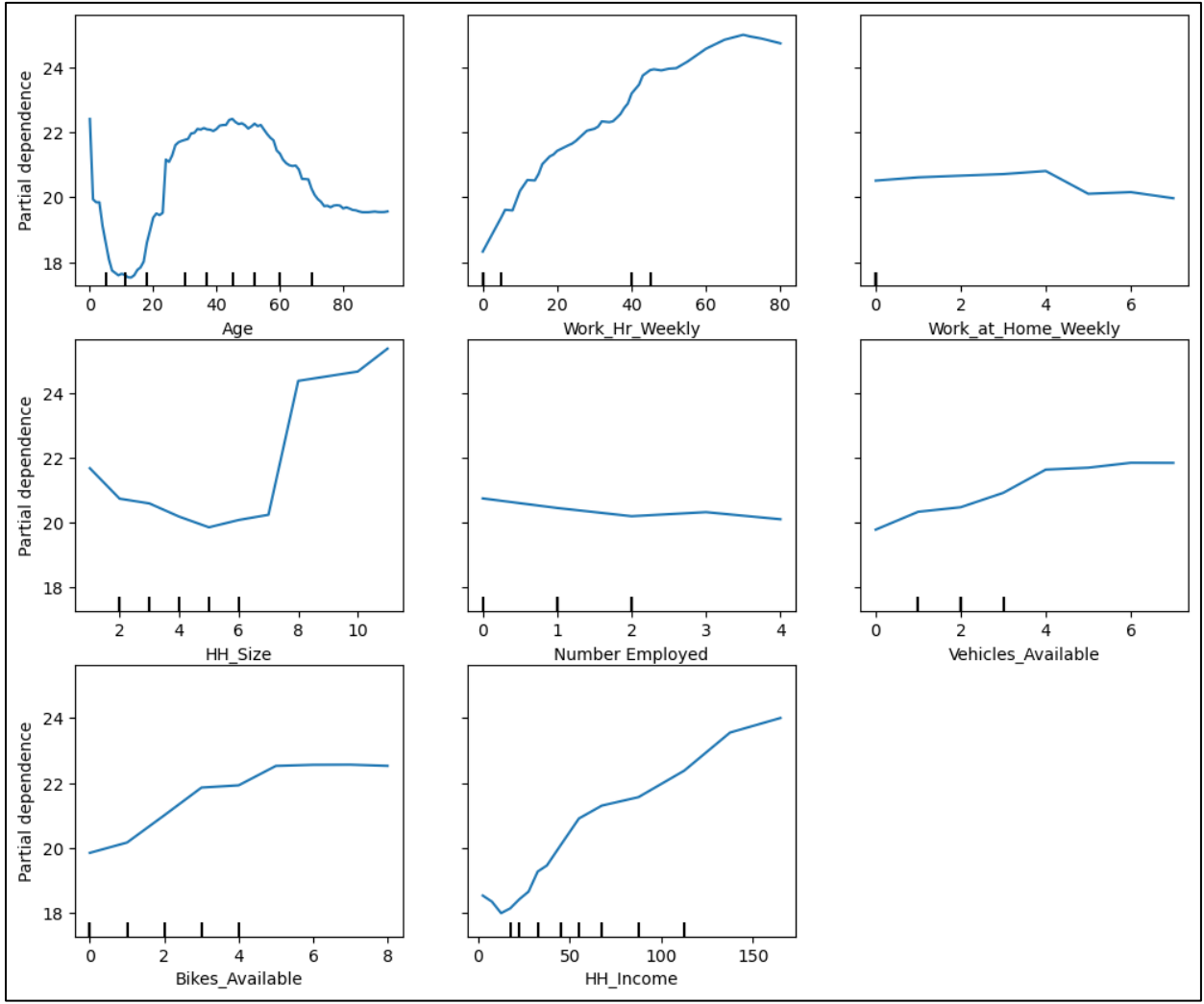

**Figure A2:** Relationship between VMT_Person and non-built environment continuous explanatory variables in 2006.

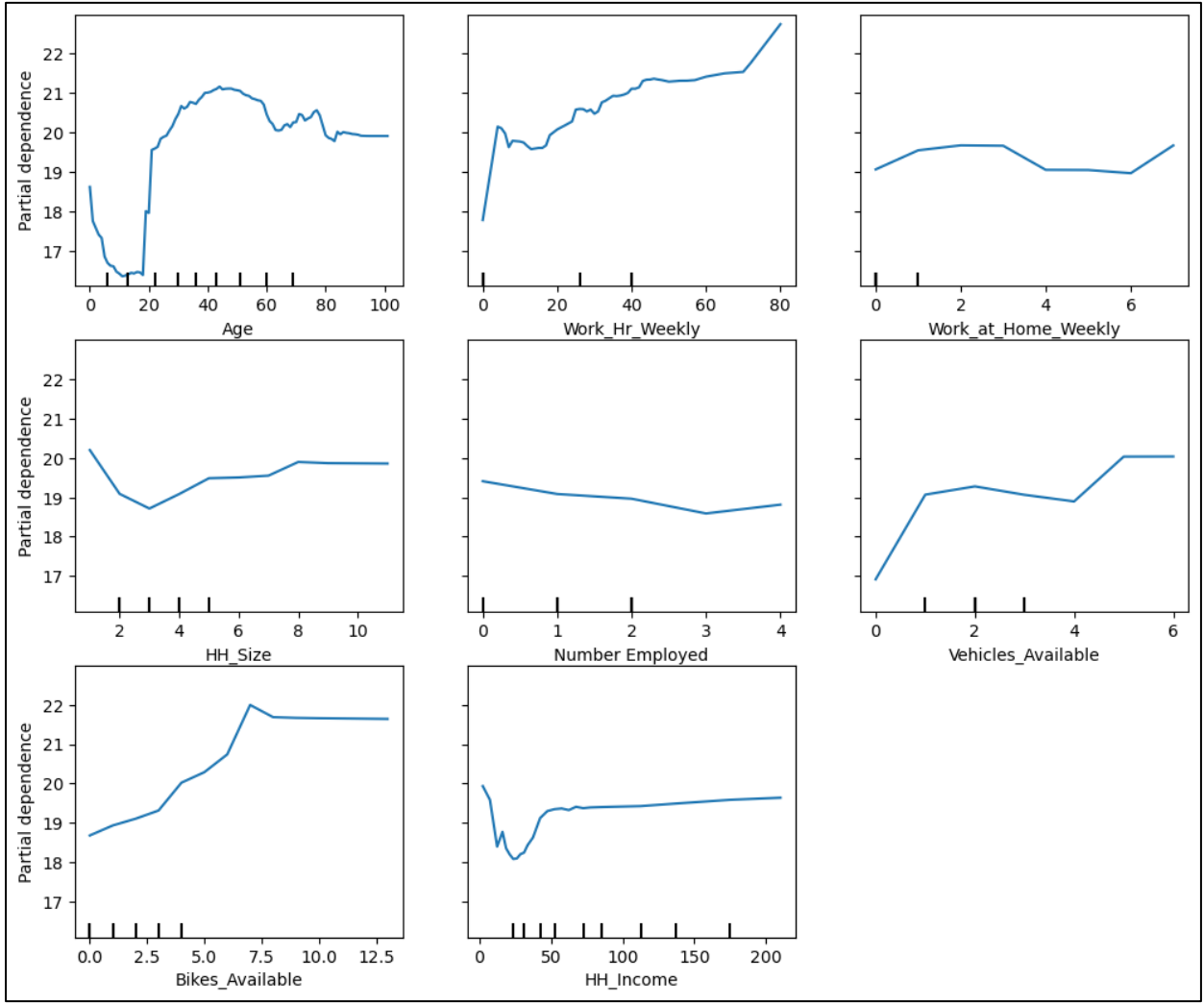

**Figure A3:** Relationship between VMT_Person and non-built environment continuous explanatory variables in 2017.

# Appendix B

**Table B1:** Fit statistics of the multilevel models

| | **Model for 1997** | | | |
|---|---|---|---|---|
| **Criteria** | **Model 1**[1] | **Model 2**[2] | **Model 3**[3] | **Model 4**[4] |
| AIC | 43325.2 | 42562.1 | 42381.1 | 42364.6 |
| BIC | 43350.8 | 42690.1 | 42547.4 | 42511.7 |
| logLik | -21658.6 | -21261.1 | -21164.5 | -21159.3 |
| Marginal $R^2$ | 0.00 | 0.14 | 0.24 | 0.25 |
| Conditional $R^2$ | 0.42 | 0.55 | 0.55 | 0.55 |
| Individual (Residual) | 26.07 | 23.11 | 23.09 | 23.10 |
| Household (Level 2) | 16.89 | 17.55 | 17.59 | 17.69 |
| TAZ (Level 3) | 14.31 | 13.86 | 8.13 | 7.10 |
| | **Model for 2006** | | | |
| AIC | 26673.9 | 26020.2 | 25890.1 | 25857.0 |
| BIC | 26697.9 | 26170.3 | 26076.2 | 26037.2 |
| logLik | -13333.0 | -12985.1 | -12914.1 | -12898.5 |
| Marginal $R^2$ | 0.00 | 0.19 | 0.26 | 0.27 |
| Conditional $R^2$ | 0.37 | 0.54 | 0.54 | 0.54 |
| Individual (Residual) | 16.95 | 14.74 | 14.75 | 14.75 |
| Household (Level 2) | 11.72 | 12.16 | 11.36 | 11.22 |
| TAZ (Level 3) | 5.82 | 3.76 | 0.00 | 0.00 |
| | **Model for 2017** | | | |
| AIC | 54399.9 | 53299.5 | 53090.6 | 53049.2 |
| BIC | 54426.9 | 53481.4 | 53312.9 | 53278.2 |
| logLik | -27196.0 | -26622.8 | -26512.3 | -26490.6 |
| Marginal $R^2$ | 0.00 | 0.14 | 0.20 | 0.21 |
| Conditional $R^2$ | 0.27 | 0.46 | 0.46 | 0.46 |
| Individual (Residual) | 16.59 | 14.51 | 14.49 | 14.48 |
| Household (Level 2) | 8.86 | 9.81 | 9.85 | 9.78 |
| TAZ (Level 3) | 4.97 | 5.00 | 1.18 | 0.63 |

[1]Model 1: Null Model
[2]Model 2: Model with individual (Level 1) and household (Level 2) characteristics as predictors, without any built environment-related predictors
[3]Model 3: Model 2 with the addition of all built environment-related predictors, without considering their non-linear effects
[4]Model 4: Model 2 with the addition of all built environment-related predictors, including their non-linear effects

Ewing, R., & Cervero, R. (2010). Travel and the Built Environment: A Meta-Analysis. *Journal of the American Planning Association*, *76*(3), 265–294. https://doi.org/10.1080/01944361003766766

Ewing, R., & Cervero, R. (2017). "Does Compact Development Make People Drive Less?" The Answer Is Yes. *Journal of the American Planning Association*, *83*(1), 19–25. https://doi.org/10.1080/01944363.2016.1245112

Gao, J., Kamphuis, C. B. M., Ettema, D., & Helbich, M. (2019). Longitudinal changes in transport-related and recreational walking: The role of life events. *Transportation Research Part D: Transport and Environment*, *77*, 243–251. https://doi.org/10.1016/j.trd.2019.11.006

Grunfelder, J., & Nielsen, T. S. (2012). Commuting behaviour and urban form: A longitudinal study of a polycentric urban region in Denmark. *Geografisk Tidsskrift-Danish Journal of Geography*, *112*(1), 2–14. https://doi.org/10.1080/00167223.2012.707806

Handy, S. (2017). Thoughts on the Meaning of Mark Stevens's Meta-Analysis. *Journal of the American Planning Association*, *83*(1), 26–28. https://doi.org/10.1080/01944363.2016.1246379

Handy, S., Cao, X., & Mokhtarian, P. (2005). Correlation or causality between the built environment and travel behavior? Evidence from Northern California. *Transportation Research Part D: Transport and Environment*, *10*(6), 427–444. https://doi.org/10.1016/j.trd.2005.05.002

Handy, S., Cao, X., & Mokhtarian, P. L. (2006). Self-Selection in the Relationship between the Built Environment and Walking: Empirical Evidence from Northern California. *Journal of the American Planning Association*, *72*(1), 55–74. https://doi.org/10.1080/01944360608976724

Hatami, F., Rahman, Md. M., Nikparvar, B., & Thill, J.-C. (2023). Non-Linear Associations Between the Urban Built Environment and Commuting Modal Split: A Random Forest Approach and SHAP Evaluation. *IEEE Access*, *11*, 12649–12662. IEEE Access. https://doi.org/10.1109/ACCESS.2023.3241627

Heres, D. R., & Niemeier, D. A. (2017). The Past and Future of Research on the Link Between Compact Development and Driving: Comment on "Does Compact Development Make People Drive Less?" *Journal of the American Planning Association*, *83*(2), 145–148. https://doi.org/10.1080/01944363.2017.1279949

Hu, X., Cao, Y., Peng, T., Gao, R., & Dai, G. (2021). Nonlinear Influence Model of Built Environment of Residential Area on Electric Vehicle Miles Traveled. *World Electric Vehicle Journal*, *12*(4), Article 4. https://doi.org/10.3390/wevj12040247

Kamruzzaman, Md., Washington, S., Baker, D., Brown, W., Giles-Corti, B., & Turrell, G. (2016). Built environment impacts on walking for transport in Brisbane, Australia. *Transportation*, *43*(1), 53–77. https://doi.org/10.1007/s11116-014-9563-0

Knaap, G.-J., Avin, U., & Fang, L. (2017). Driving and Compact Growth: A Careful Look in the Rearview Mirror. *Journal of the American Planning Association*, *83*(1), 32–35. https://doi.org/10.1080/01944363.2017.1251276

Krizek, K. J. (2000). Pretest-Posttest Strategy for Researching Neighborhood-Scale Urban Form and Travel Behavior. *Transportation Research Record*, *1722*(1), 48–55. https://doi.org/10.3141/1722-06

Krizek, K. J. (2003). Residential Relocation and Changes in Urban Travel: Does Neighborhood-Scale Urban Form Matter? *Journal of the American Planning Association*, *69*(3), 265–281. https://doi.org/10.1080/01944360308978019
38